\documentclass{mem}
\usepackage{natbib}\usepackage{txfonts}\usepackage{balance}
\usepackage{graphicx}
\usepackage[breaklinks,dvipdfm]{hyperref}
\idline{75}{282}
\begin{document}
\def\teff{$T\rm_{eff }$}
\def\kms{$\mathrm {km s}^{-1}$}

\title{
The Second Parameter Problem(s)
}

   \subtitle{}

\author{Aaron Dotter\inst{1}}

  \offprints{A.\ Dotter}

\institute{
Research School of Astronomy and Astrophysics
The Australian National University
Canberra ACT 2611
\email{aaron.dotter@gmail.com}
}

\authorrunning{Dotter}

\titlerunning{The Second Parameter Problem(s)}

\abstract{
The Second Parameter (2ndP) Problem recognizes the remarkable role played by 
horizontal branch (HB) morphology in the development of our understanding of 
globular clusters, and the Galaxy, over the last 50 years.  I will describe the 
historical development of the 2ndP and discuss recent advances that are finally
providing some answers.  I will discuss how the controversies surrounding the 
nature of the 2ndP can be reconciled if we acknowledge that there are actually 
two distinct problems with entirely different solutions.

}
\maketitle{}

\section{Introduction}

The Second Parameter (2ndP) Problem arose in the 1960's when astronomers
were first able to assemble a considerable sample of globular cluster (GC)
color-magnitude diagrams (CMDs). The CMDs revealed that horizontal branch (HB)
morphology was mainly driven by the metallicity, but that metallicity alone
was insufficient to explain the observed diversity of HB morphology. Early
references to an `other' or `second' parameter can be found in \citet{SW67}
and \citet{vdB67}.

\citet{SZ78} tied the 2ndP problem to the larger problem of formation of the 
Galaxy with the argument that the appearance of metal-poor GCs with red HB
morphologies in the outer Galactic halo was indicative of it having formed
by the gradual accretion of so-called `protogalactic fragments.' Thus, 
\citet{SZ78} linked the 2ndP with age. While the `hierarchical' 
formation scenario proposed by \citet{SZ78} has gained wide acceptance as a 
galaxy formation model over the years, the suggestion that the 2ndP is age
remains controversial.

The remainder of this paper will discuss the reasons for the controversy 
surrounding the identity of the 2ndP and attempt to explain the root
cause(s).

\section{An historical view of globular cluster ages}\label{ages}
At least since the seminal paper by \citet{SZ78}, the ages of the Galactic GCs
have been the subject of many studies; for a review see, for example, the 
introduction to \citet{D10}. Here I focus on those studies that target large
samples of GCs.  The chief interest of these studies is
the question of whether or not the Galactic GC population exhibits an age
gradient as a function of metallicity, Galactocentric radius, or some other
parameter.

\begin{figure*}
\includegraphics[width=0.5\textwidth]{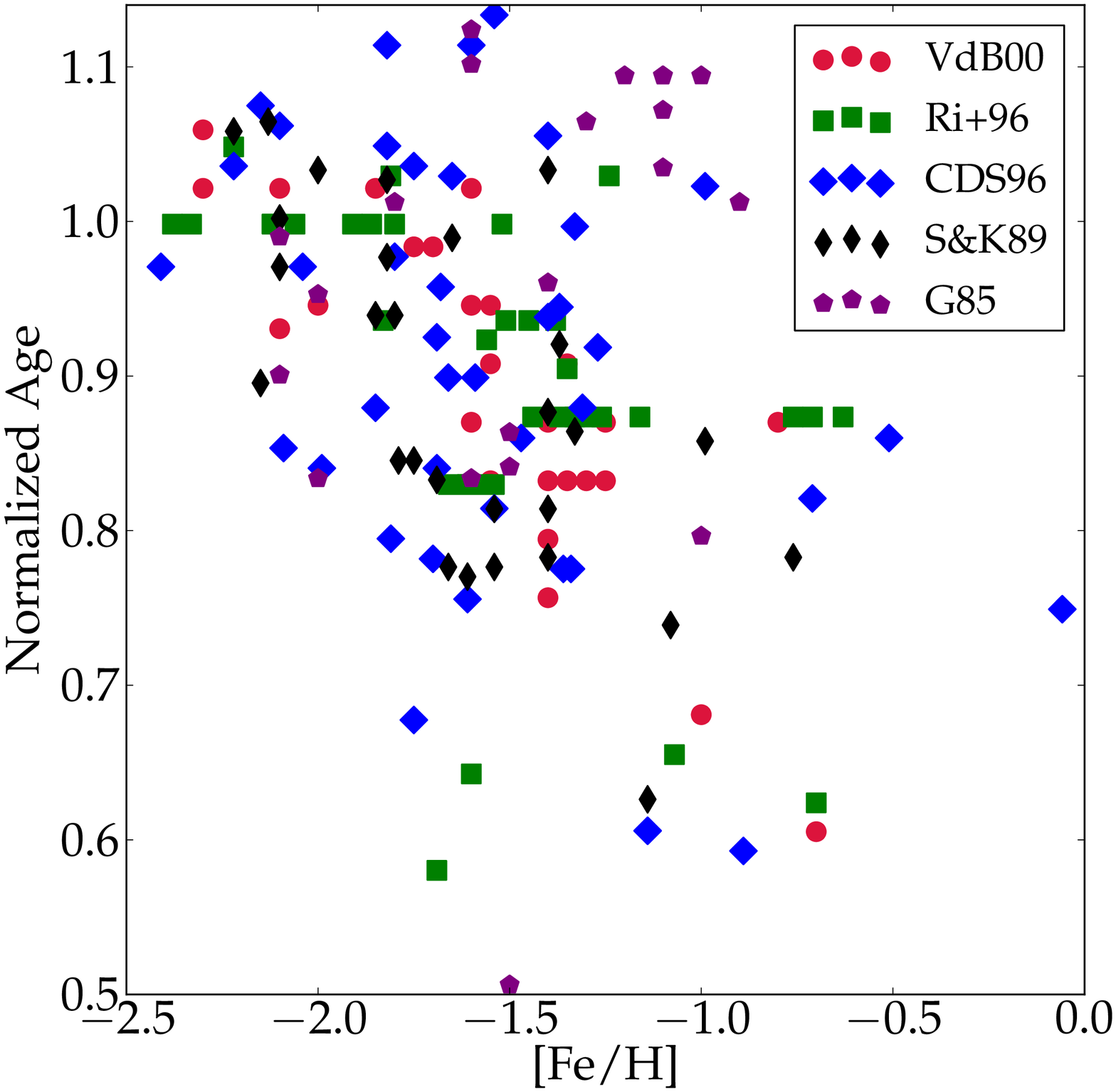}
\includegraphics[width=0.5\textwidth]{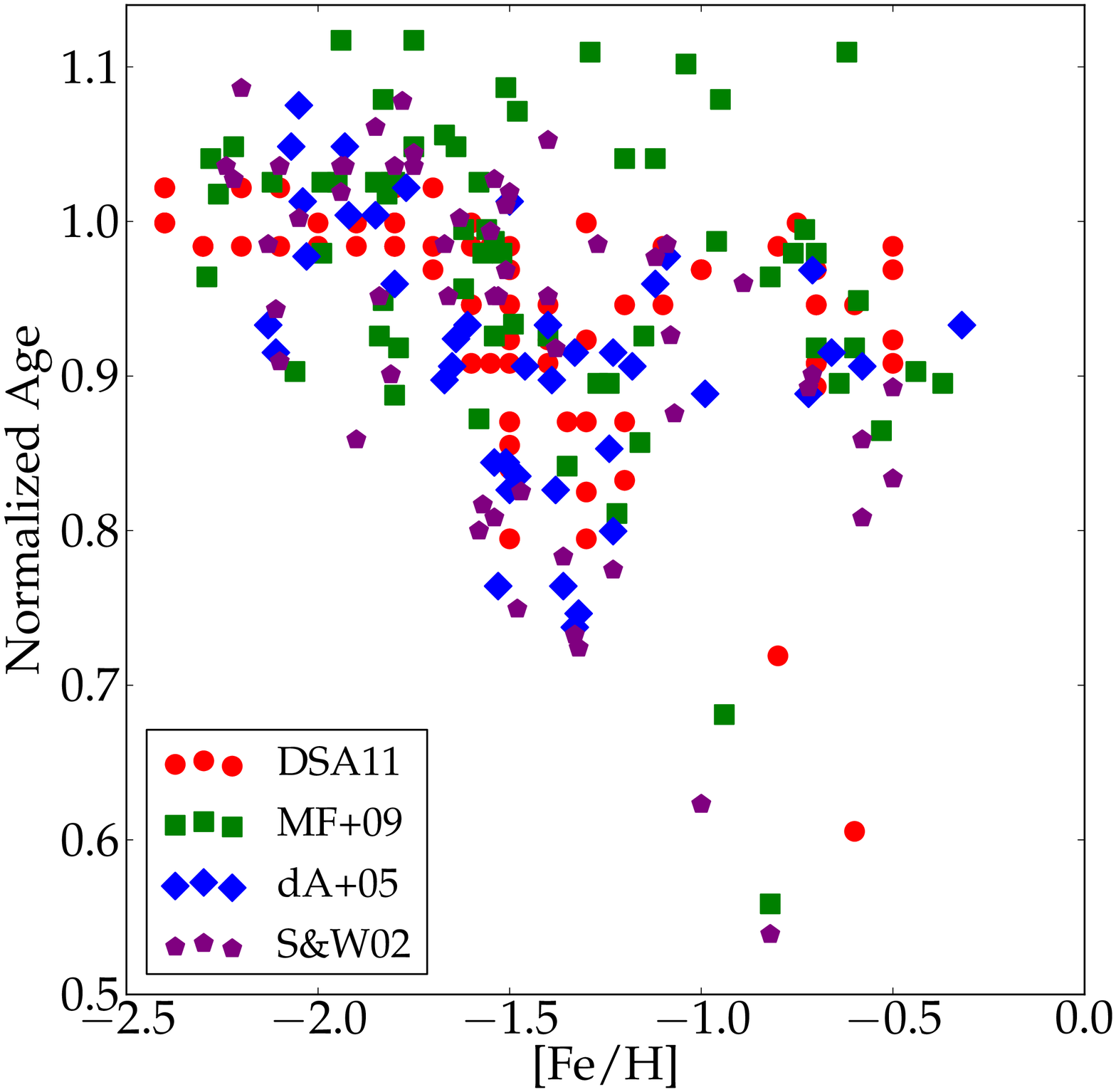}
\caption{\footnotesize Normalized ages for a total of 9 studies published 
between 1985 and the present. References: G85=\citet{G85}, SK89=\citet{SK89},
CDS96=\citet{CDS96}, Ri+96=\citet{Ri96}, VdB00=\citet{VdB00}, 
S\&W02=\citet{SW02}, dA+05=\citet{dA05}, MF+09=\citet{MF09}, DSA11=\citet{DSA11}
}
\label{fig:norm}
\end{figure*}

Figure \ref{fig:norm} displays a comparison of GC age studies published between
1985 and the present. The studies considered here are not necessarily complete,
 but should be representative. The left panel of Figure \ref{fig:norm} shows 
studies 
published between the years 1985 and 2000. The right panel shows studies 
published between the year 2001 and the present. In order to put the ages
on a comparable scale a Normalized Age has been defined as the age relative
to the average age of the metal-poor GCs in each sample. Here `metal-poor'
means all GCs with [Fe/H] $\leq -1.7$.\footnote{The Normalized Age for a given
study relies only on the information provided within that study. That is, it
uses both the adopted ages and metallicities of that study.}
This step is necessary to compare in
particular the older studies since the typical age scale of GCs decreased
substantially from the 1980's to the present as a result of improvements to
stellar evolution models.
The range of plotted Normalized Ages and [Fe/H] is the same in both panels.

Figure \ref{fig:norm} reveals that the earlier studies (plotted in the left 
panel) show some evidence of an age-metallicity relation but there is
considerable scatter in age at a fixed [Fe/H]. The later studies (plotted
in the right panel) shows much tighter agreement among the different studies
\emph{and} all of them indicate the presence of a bifurcation in the 
age-metallicity relation for GCs with [Fe/H] $\geq -2$ \citep[see discussions
in, e.g.,][]{MF09,DSA11}. It is my contention that the majority of the 
improved agreement among the later studies is due to the improved quality
and homogeneity of the observations afforded by the Hubble Space Telescope
\citep{Pi02,Sa07}.

\begin{figure*}[ht]
\includegraphics[width=1.0\textwidth]{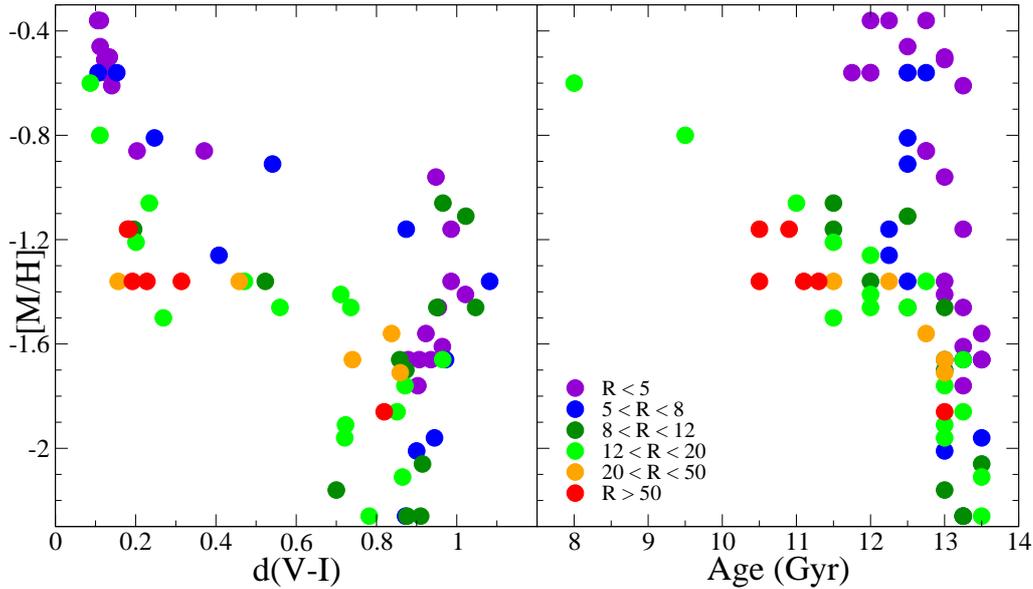}
\caption{\footnotesize The HB morphology-metallicity and age-metallicity
relations from \citet{DSA11} plotted side-by-side to demonstrate the 
correspondence between age and HB morphology \emph{in this diagram}.}
\label{fig:HBAMR}
\end{figure*}
 
To finish the discussion of GC age and HB morphology, I present the 
side-by-side comparison of the HB morphology-metallicity and age-metallicity
diagrams based on \citet{DSA11} in Figure \ref{fig:HBAMR}. Figure 
\ref{fig:HBAMR} is my version of the famous diagram by \citet[][their 
Figure 10]{SZ78}; it shows the relationship between HB morphology and age
for several bins of Galactocentric radius \emph{in this diagram}. It is 
important to emphasize the final three words in the previous sentence
because the way in which in HB morphology is represented has important 
consequences, as will be discussed in the following sections.

\section{Whence the `faint blue tails?'}\label{blue}
HB morphology is often complex and cannot be condensed into a single number
without (sometimes considerable) loss of information. The 
studies mentioned in Section \ref{ages} all rely on HB metrics that favor
the central tendency of the distribution. Furthermore, these studies rely on 
optical
CMDs that favor the redder HB stars. Clearly, such emphasis focuses the outcome
of these studies to parameters that most strongly influence the center or
red edge of the distribution.  What about the blue edge of the distribution?

\citet{FP93} demonstrated that a class of HB metrics that target the blue
end of the distribution, the so-called `faint blue tails,' reveal a tight
correlation with the absolute magnitude of the GC. Since absolute magnitude 
and mass are tightly linked quantities, the total mass of the cluster must
be related to the prevalence of hot, blue HB stars. Similar results were 
later obtained by, e.g., \citet{RB06} and \citet{G10}. It is important to
note that \citet{FP93}, \citet{RB06}, and \citet{G10} all rely on optical
CMDs for their analyses.  Yet, even using data that undervalues the hottest
HB stars, it is still possible to define an HB morphology metric that is
sensitive to them.

At the close of Section \ref{ages}, the point was made that the way in 
which an HB morphology metric is defined weighs heavily on the outcome 
of a study which adopts that metric. However, it also becomes clear that 
different parameters influence the distribution of HB stars in different 
ways. While the traditional HB metrics seem to correlate with age, at least
in some studies \citep{G10,D10}, the blue-sensitive HB metrics used by 
\citet{FP93}, \citet{RB06}, and \citet{G10} correlate with absolute magnitude.
That absolute magnitude is in some way connected with the strength of the
second generation GC stars, and therefore initial He content, is discussed
by, e.g., \citet{RB06} and \citet{G10}.

The comparison of different HB metrics indicates that they probe different
aspects of the distribution of HB stars.  See, for example, the discussions
in Section 2.2 (Figures 7 and 8) of \citet{G10} and Section 3.3 (Figure 2)
of \citet{D10}. As already mentioned, this leads to a different outcome,
even when using the same data set, if only one HB metric is considered.

When weighing the relative contributions of different candidates 
for the 2ndP, it therefore becomes necessary to consider more than one
HB metric. Studies by \citet{G10} and A.\ Milone (this volume) deserve
special attention in this regard.

The difficulty encountered in using optical CMDs can be largely removed
by using combined UV-optical CMDs. An example of the UV-optical studies
is that of NGC 2808 by \citet{Da11}, who study the complex HB of this
peculiar, massive GC using WFPC2 data. Their Figure 3 shows the NGC 2808
HB (and also that of M\,80) in the $F160BW-F555W$ CMD, which spans more 
than 6 magnitudes in the color dimension while it is nearly flat in 
$F160BW$ magnitude, except for a few of the reddest HB stars.

\section{Discussion}
In the language of 
\emph{descriptive statistics},\footnote{See http://en.wikipedia.org/wiki/Descriptive\_statistics} some HB metrics 
probe the central tendency of the distribution while others probe the extremes. 
It is easy to imagine that one parameter (e.g., metallicity or age) can drive 
the central tendency, while another (e.g., the range of initial He content) can 
drive the dispersion and/or the shape of the distribution.

\citet{FN81} wrote
\begin{quote}
at least two parameters in addition to [Fe/H] are probably necessary to explain
the observed HB morphologies. One of these must be a parameter that varies 
within clusters (i.e., a non-global parameter) which the second varies from 
cluster to cluster (i.e., a global parameter). Almost all discussion to this
time has centered on the global parameters.
\end{quote}
The meaning of `global' in this context pertains all stars within a single GC. 
We can identify parameters, such as the bulk metallicity ([Fe/H]) and age, as 
global in the sense that they do not vary substantially within a (normal) GC; other 
parameters, such as the degree of internal pollution, can be associated 
with the sense that a range of these parameters exists within a single GC.
Moreover, the degree to which a non-global parameter varies within one GC
may differ considerably with the degree to which the same parameter varies
within another GC.

I close the discussion with the suggestion that the global parameters are, 
in a sense, connected with the central tendency while the non-global 
parameters are connected with the shape and dispersion of the distribution.

\begin{acknowledgements}
I wish to thank the organizers for an excellent meeting and Ata Sarajedini 
for inspiring my interest in the Second Parameter.
\end{acknowledgements}

\bibliographystyle{aa}

\end{document}